\documentclass[PRL,twocolumn,showpacs,amsmath,amssymb,superscriptaddress]{revtex4}

\newcommand{\vague}[1]{`#1'}                                               %:text:
\newcommand{\spcpnct}{\;}                              % space before punctation in equation
\newcommand{\period}{{\mbox{\spcpnct.}\relax}}         % period at the end of equation
\newcommand{\commae}{{\mbox{\spcpnct,}\relax}}         % comma at the end of equation
\newcommand{\comma}{{\mbox{\spcpnct,\quad}\relax}}     % comma between equations
\newcommand{\sect}[1]{\section*{#1}}                   % section headding

\newcommand{\srh}{\frac1{\sqrt2}}                      % just a shortcut

\newcommand{\ctr}{\cdot}                               % contraction
\newcommand{\spr}{\cdot}                               % scalar product
\newcommand{\tens}[1]{{\boldsymbol{#1}}}               % tensor wrapper            %:ex: \tens{a}
\newcommand{\cv}[1]{{\tens{\partial}}_{#1}}            % coordinate vectors        %:ex: \cv{x_\mu} \quad \cv{\psi_j}
\newcommand{\grad}{{\mathbf{d}}}                       % gradient and external derivative
\newcommand{\lied}{\pounds}                            % Lie derivative
\newcommand{\covd}{{\tens{\nabla}}}                    % covariant derivative
\newcommand{\Du}[1]{{D_{#1}}}                          % covariant derivative in \muv direction
\newcommand{\Db}[1]{{\bar{D}_{#1}}}                    % covariant derivative in \mbv direction
\newcommand{\mtrc}{{\tens{g}}}                         % metric
\newcommand{\pKY}{{\tens{h}}}                          % principal CCKY
\newcommand{\PKY}{{\tens{\check h}}}                   % principal CCKY as an operator
\newcommand{\PKV}{{\tens{\xi}}}                        % principal KV
\newcommand{\env}[1]{{\tens{e}_{#1}}}                  % vector in $x_\mu$ direction                        %:ex: \env{\mu}
\newcommand{\enf}[1]{{\tens{e}^{#1}}}                  % 1-form in $x_\mu$ direction                        %:ex: \enf{\mu}
\newcommand{\ehv}[1]{{\hat{\tens{e}}_{#1}}}            % vector in $\mu$th 2-plane in Killing sector        %:ex: \ehv{\mu}
\newcommand{\ehf}[1]{{\hat{\tens{e}}^{#1}}}            % 1-form in $\mu$th 2-plane in Killing sector        %:ex: \ehf{\mu}
\newcommand{\muv}[1]{{\tens{m}_{#1}}}                  % null vector in $\mu$th 2-plane                     %:ex: \muv{\mu}
\newcommand{\mbv}[1]{{\bar{\tens{m}}_{#1}}}            % barred null vector in $\mu$th 2-plane              %:ex: \mbv{\mu}
\newcommand{\cxv}[1]{{\tens{\epsilon}_{#1}}}           % vector in $x_\mu$ direction                        %:ex: \cxv{\mu}=\cv{x_\mu}
\newcommand{\cxf}[1]{{\tens{\epsilon}^{#1}}}           % 1-form in $x_\mu$ direction                        %:ex: \cxf{\mu}=\grad{x_\mu}
\newcommand{\chv}[1]{{\hat{\tens{\epsilon}}_{#1}}}     % the i-th Killing vector                            %:ex: \chv{i}=\cv{\psi_i}
\newcommand{\chf}[1]{{\hat{\tens{\epsilon}}^{#1}}}     % gradient of the i-th Killing coordinate            %:ex: \chf{i}=\grad{\psi_i}
\begin{document}
\title{Hidden Symmetries of Higher Dimensional Black Holes and
Uniqueness of the Kerr-NUT-(A)dS spacetime}

\author{Pavel Krtou\v{s}}

\email{Pavel.Krtous@utf.mff.cuni.cz}

\affiliation{Institute of Theoretical Physics, Faculty of Mathematics and Physics, Charles University in Prague,
V~Hole\v{s}ovi\v{c}k\'ach 2, Prague, Czech Republic}

\author{Valeri P. Frolov}

\email{frolov@phys.ualberta.ca}

\affiliation{Theoretical Physics Institute,\\ University of Alberta, Edmonton,
Alberta, Canada T6G 2G7}

\author{David Kubiz\v n\'ak}

\email{kubiznak@phys.ualberta.ca}

\affiliation{Theoretical Physics Institute,\\ University of Alberta, Edmonton,
Alberta, Canada T6G 2G7}

\affiliation{Institute of Theoretical Physics, Faculty of Mathematics and Physics, Charles University in Prague,
V~Hole\v{s}ovi\v{c}k\'ach 2, Prague, Czech Republic}

\date{April 29, 2008}   % version 2.00 arXiv version

\begin{abstract}
We prove that the most general solution of the Einstein equations
with the cosmological constant which admits a principal conformal
Killing--Yano tensor is the Kerr-NUT-(A)dS metric.  Even when the
Einstein equations are not imposed, any spacetime admitting such 
hidden symmetry can be written in a canonical form which guarantees the
following properties: it is of the Petrov type D, it allows the
separation of variables for the Hamilton--Jacobi, Klein--Gordon, and
Dirac equations, the geodesic motion in such a spacetime is
completely integrable.  These results naturally generalize the
results obtained earlier in four dimensions.
\end{abstract}

\pacs{04.50.-h, 04.50.Gh, 04.70.Bw, 04.20.Jb \hfill  Alberta-Thy-07-08}
% 04.50.-h  Higher-dimensional gravity and other theories of gravity
% 04.50.Gh  Higher-dimensional black holes, black strings, and related objects
% 04.70.Bw  Classical black holes
% 04.20.Jb  Exact solutions

\maketitle

\sect{Introduction} Higher-dimensional black holes which might play the
role of natural probes of extra dimensions are being discussed
intensively at present (see, e.g., the review \cite{EmparanReall:2008} and
references therein). Recently, the subject of hidden symmetries of
higher-dimensional black hole metrics has become of high
interest. By studying the hidden symmetries it was demonstrated that
higher-dimensional black holes are in many aspects similar to their
four-dimensional `cousins'. 

Explicit spacetime symmetries are represented by Killing vectors. Hidden
symmetries are related to generalizations of this concept.
One of the most important of these generalizations is the hidden 
symmetry encoded in the \emph{principal conformal Killing--Yano (CKY) tensor} 
\cite{Tachibana,KrtousEtal:2007a}. 
%\cite{Tachibana:1968,Tachibana:1969,KrtousEtal:2007a}. 
It was demonstrated that the Myers--Perry metric
\cite{MyersPerry:1986}, describing the higher-dimensional rotating
black hole, as well as its generalization, the Kerr-NUT-(A)dS metric
\cite{ChenLuPope:2006} which includes the NUT parameters and the
cosmological constant, admit such a tensor \cite{FrolovKubiznak}. 

The principal CKY tensor generates hidden symmetries 
which are responsible for the existence of non-reducible quadratic in momenta
conserved integrals of motion for geodesic motion. In consequence, it was shown
that geodesic equations in the Kerr-NUT-(A)dS spacetime are
completely integrable \cite{PageEtal,KrtousEtal:2007a}. 
Moreover, the Hamilton--Jacobi, Klein--Gordon, 
and Dirac equations are separable in this
spacetime \cite{FrolovEtal:2007,SergyeyevKrtous:2007,OotaYasui:2007}. This spacetime 
is of the algebraic type D \cite{HamamotoEtal:2007,PravdaEtal:2007} 
of the higher-dimensional generalization of the Petrov classification
\cite{ColeyEtal}
%\cite{ColeyMilsonPravdaPravdova:2004a,MilsonColeyPravdaPravdova:2005,Coley:2008}, 
and can be presented in the generalized Kerr--Schild form \cite{ChenLu:2007}. All these
remarkable properties make higher-dimensional black hole solutions very
similar to the 4D black holes   (see, e.g.,
\cite{FrolovKubiznak:2008a,Frolov:2007} for a review).

The study of 4D metrics admitting the principal CKY tensor showed
that the corresponding  class of solutions of the Einstein equations with
the cosmological constant reduces to the Kerr-NUT-(A)dS spacetime
\cite{DietzRudigerTaxiarchis}. 
%\cite{DietzRudiger:1981,Taxiarchis:1985}. 
%\cite{DietzRudiger:1981,DietzRudiger:1982}.
The aim of this paper is to demonstrate
that in the higher-dimensional gravity the situation is similar.
Namely, we shall prove the following results: The metric of any 
higher-dimensional spacetime which admits a principal CKY tensor 
can be put into a special \emph{canonical form} 
which guarantees the following properties: 
(1) it is of the algebraic type~D, 
(2) it allows a separation of variables for
the Hamilton--Jacobi, Klein--Gordon, and Dirac equations, and 
(3) the geodesic motion in such a spacetime is completely integrable.
When the Einstein equations with the cosmological constant are imposed
the canonical form becomes the Kerr-NUT-(A)dS metric \cite{ChenLuPope:2006}.

Similar results were proved recently 
for spacetimes with a principal CKY tensor obeying special additional restrictions
%\cite{HouriEtal:2007,HouriEtal:2008}
\cite{HouriEtal}.
The conditions imposed, however, might have seemed very restrictive. 
In this paper we demonstrate that it is not so. Namely, 
we demonstrate that these restrictions are not necessary and that they in fact 
automatically follow from the properties of 
the principal CKY tensor. 
In particular, this means that all the results
\cite{PageEtal,KrtousEtal:2007a,FrolovEtal:2007,SergyeyevKrtous:2007,OotaYasui:2007,HamamotoEtal:2007} (proved for the canonical form of the metric) immediately follow from the very existence of the principal 
CKY tensor. It also means that one cannot hope to arrive at more general spacetimes by relaxing the 
additional assumptions of \cite{HouriEtal}.

Here we present only a sketch of the proof of our results and 
for simplicity restrict to an even dimension ${D=2n}$.
All technical details, including the odd-dimensional case, 
will be discussed in \cite{proofpaper}.

%For simplicity, we present our results only in an even dimension ${D=2n}$.
%The analogous result in the odd dimensional case will be presented in 
%\cite{proofpaper}, where all the technical
%details of our construction will also be presented.

%To simplify the presentation we restrict ourselves to the case of even dimensions ${D=2n}$.
%The generalization to the odd-dimensional case is straightforward 
%and will be discussed in \cite{proofpaper}, where also all the technical
%details of our construction will be presented.

\sect{Definitions and assumptions} 
A principal conformal Killing--Yano tensor is defined as  a closed
non-degenerate 2-form ${\pKY}$ obeying the following equation
\begin{equation}\label{CKYeq}
\nabla_c h_{ab}=g_{ca}\,\xi_{b}-g_{cb}\,\xi_{a}\, .
\end{equation} 
This equation implies
\begin{equation}\label{xidef}
\nabla_{[a}h_{bc]}=0\comma \xi_a=\frac1{D-1} \nabla^n h_{na}\period
\end{equation}
In what follows we shall assume that ${\PKV\ne 0}$. The case when
${\PKV=0}$, and hence $\pKY$ is covariantly constant, requires a
special consideration. The condition of non-degeneracy in an even dimension 
means that the skew symmetric matrix $h_{ab}$ in the ${D=2n}$ dimensional
spacetime has the matrix rank ${2n}$. 

As any 2-form on a metric space, the principal CKY tensor
${\pKY}$ determines an orthonormal 
\cite{endnote1}
\emph{Darboux basis} which simultaneously diagonalizes the metric ${\mtrc}$ and 
\vague{skew-diagonalizes} ${\pKY}$. Namely, there exists a frame
${\{\enf{\mu},\ehf{\mu}\}}$, ${\mu=1,\dots,n}$, of 1-forms in which
the metric and the principal CKY tensor are
\begin{equation}\label{ghdiag}
 \mtrc=\sum_\mu\bigl(\enf{\mu}\enf{\mu}+\ehf{\mu}\ehf{\mu}\bigr)\comma
 %+\epsilon\ehf{\mu}_{\epsilon}\ehf{\mu}_{\epsilon}
 %\label{gdiag}\commae\\
 \pKY=\sum_\mu x_\mu \enf{\mu}\wedge\ehf{\mu}\period
\end{equation}
(We do not use the summation convention for Greek indices---all
sums over them are indicated explicitly and run in the range ${1,\dots,n}$
unless stated otherwise.)

We denote by $\PKY$ the operator with components $h^{a}{}_{b}$ and by 
`${\ctr}$' a symbol for contraction. For example, $\PKY\ctr\PKY\ctr\tens{v}$ 
denotes a vector with the
components ${h^{a}{}_{b}\,h^{b}{}_{c}\,v^c}$.
The operator $\PKY$ is antisymmetric 
with respect to the metric scalar product. This means that the operator
${-\PKY{}^2=-\PKY\ctr\PKY}$ is a non-negative definite symmetric 
operator, and its eigenvectors are given by vectors 
${\{\env{\mu},\ehv{\mu}\}}$ of the vector frame dual to the Darboux basis
\begin{equation}\label{eveqh2}
-\PKY{}^2\ctr \env{\mu} = x_{\mu}^2\,\env{\mu}\comma
-\PKY{}^2\ctr \ehv{\mu} = x_{\mu}^2\,\ehv{\mu}\period
\end{equation}
As discussed below, we assume that the eigenspaces corresponding
to each eigenvalue $x^2_{\mu}>0$ are two-dimensional and we call them
the {\em Killing--Yano (KY) 2-planes}. 

It is convenient to introduce also a basis of complex
null eigenvectors ${\{\muv{\mu},\mbv{\mu}\}}$ which obey the relations
\begin{equation}\label{eveqs}
\PKY\ctr\muv{\mu}=-ix_{\mu} \muv{\mu}\comma
\PKY\ctr\mbv{\mu}= ix_{\mu} \mbv{\mu}\commae
\end{equation}
with bar denoting the complex conjugation.
These complex null vectors satisfy the normalization
\begin{equation}\label{mnorm}
\muv{\mu}\spr\muv{\nu}=\mbv{\mu}\spr\mbv{\nu}=0\comma
\muv{\mu}\spr\mbv{\nu}=\delta_{\mu\nu}\commae
\end{equation}
and they are connected with vectors 
${\{\env{\mu},\ehv{\mu}\}}$ from Eq.~\eqref{eveqh2} as follows:
\begin{equation}\label{mdef}
  \muv{\mu}=\srh(\ehv{\mu}+i\,\env{\mu})\comma
  \mbv{\mu}=\srh(\ehv{\mu}-i\env{\mu})\period
\end{equation}

\sect{Uniqueness of the Kerr-NUT-(A)dS metric} 
Passing from a local description of the principal 
CKY tensor $\pKY$ at a chosen
point to the description of a spacetime properties in some domain, we
include into the notion of the principal CKY tensor
the following requirement: the \vague{eigenvalues} ${x_\mu}$ of $\pKY$ are 
functionally independent in some spacetime domain, that is we
assume that ${x_\mu}$ are non-constant independent scalar functions
with different values at a generic point.

This also allows us to use ${x_\mu}$\!'s as coordinates. 
As a part of our result we demonstrate
that these $n$ coordinates can be upgraded by adding $n$
new coordinates $\psi_{i}$ so that the metric and the principal CKY tensor take the form
\begin{gather}
  \mtrc=\sum_\mu\Bigl[\frac1{Q_\mu}\,\grad x_\mu\grad x_\mu+
  Q_\mu \Bigl(\sum_{i=0}^{n-1} A^i_\mu \grad \psi_i\Bigr)
  \Bigl(\sum_{j=0}^{n-1} A^j_\mu \grad \psi_j\Bigr)\Bigr]\commae\notag\\
  \pKY=\grad\tens{b}\comma
  \tens{b}=\frac12\sum_{j=0}^{n-1}A^{j+1}\grad\psi_j\period
  \label{canform}
\end{gather}
Here, functions ${A^i,A^i_\mu}$ and ${U_\mu}$ are particular 
combinations of ${x_\mu}$\!'s,
\begin{equation}\label{AUdef}
\begin{gathered}
  A^{i}_\mu=\!\!
  \sum_{\substack{\nu_1,\dots,\nu_i\\\nu_1<\dots<\nu_i\\\nu_j\ne\mu}}\!
   x_{\nu_1}^2\dots x_{\nu_i}^2\comma
  A^{i}=\!\!
  \sum_{\substack{\nu_1,\dots,\nu_i\\\nu_1<\dots<\nu_i}}\!
   x_{\nu_1}^2\dots x_{\nu_i}^2\commae\\[-1ex]
  U_{\mu}=\prod_{\substack{\nu\\\nu\ne\mu}}(x_{\nu}^2-x_{\mu}^2)\commae
\end{gathered}
\end{equation}
and metric functions ${Q_\mu}$ are given by
\begin{equation}\label{Qdef}
  Q_\mu = \frac{X_\mu}{U_\mu}\comma X_\mu=X_\mu(x_\mu)\commae
\end{equation}
with ${X_\mu}$ depending only on a single coordinate ${x_\mu}$. 
For odd number of dimensions the metric \eqref{canform} contains
few extra terms and its form can be found, e.g.,
in \cite{ChenLuPope:2006,HamamotoEtal:2007}.

We call \eqref{canform}--\eqref{Qdef} the \emph{canonical} form of the metric.
It can be consider as a higher-dimensional generalization of the 
form of the metric constructed by Carter in four dimensions 
\cite{Carter}.
%\cite{Carter:1968a,Carter:1968b}.
In what follows we demonstrate that
coordinates $(x_{\mu},\psi_i)$ used in 
\eqref{canform} have a well defined geometrical meaning---determined 
completely by the principal CKY tensor. It should
be emphasized that this canonical form
follows from the existence of
the principal CKY tensor \emph{off-shell}, that is without imposing 
the Einstein equations. 
When the vacuum Einstein equations with the cosmological constant are imposed, the
metric \eqref{canform} turns out to be the Kerr-NUT-(A)dS metric
\cite{ChenLuPope:2006,HamamotoEtal:2007} for which one has
\begin{equation}\label{BHXsE}
  X_\mu = b_\mu\, x_\mu + \sum_{k=0}^{n}\, c_{k}\, x_\mu^{2k}\period
\end{equation}
The constants ${c_k}$ and ${b_\mu}$ are related to the cosmological constant,
angular momenta, mass, and NUT charges, see, e.g., \cite{ChenLuPope:2006}
for details. For ${b_\mu=0}$ we obtain
the constant curvature space.

As mentioned in Introduction, the uniqueness of the
Kerr-NUT-(A)dS metric has been already studied in
\cite{HouriEtal} where it was proved
provided the following additional assumptions: 
\begin{equation}\label{Liexihg}
  \lied_\PKV \pKY  =0 \comma  %\label{Lixih}\\
  \lied_\PKV \mtrc =0\period         %\label{Lixig}
\end{equation}
That is, the authors of \cite{HouriEtal} explicitly 
required, that ${\PKV}$ is a Killing vector and that the principal CKY tensor ${\pKY}$ does not change
along ${\PKV}$. We prove now that both of
these conditions are superfluous since they follow from the existence of the principal CKY
tensor. The proof of the first condition is presented below,
the second condition is a corollary of our explicit construction of the
canonical form \eqref{canform} of the off-shell metric.

\sect{Condition on the principal CKY tensor}
First, we concentrate on the first condition in \eqref{Liexihg}. Let
us denote ${\Du{\mu}=\covd_{\muv{\mu}}}$ and
${\Db{\mu}=\covd_{\mbv{\mu}}}$. Using \eqref{CKYeq} one has
\begin{equation}\label{Dh}
(\Du{\mu}\PKY)\ctr\muv{\nu}=(\muv{\nu}\spr\PKV)\,\muv{\mu}\period
\end{equation}
Applying ${\Du{\mu}}$ to \eqref{eveqs} and using \eqref{Dh} one obtains
\begin{equation}\label{eqn}
(\PKY+i x_{\nu}\tens{\delta})\ctr\Du{\mu}\muv{\nu}
+ i (\Du{\mu} x_{\nu})\, \muv{\nu}
+(\muv{\nu}\spr\PKV) \, \muv{\mu}=0\period
\end{equation}
By taking a scalar product of \eqref{eqn} with $\mbv{\nu}$,
using antisymmetry of ${\PKY}$ and Eq.~\eqref{eveqs} again, 
the first term cancels out. 
Considering two cases when $\nu=\mu$ and when $\nu\ne \mu$ one gets
\begin{equation}\label{Dmx}
\Du{\mu}x_{\nu}=0\quad\text{for}\; \nu\ne\mu\comma
\Du{\mu}x_{\mu}=i \muv{\mu}\spr\PKV\period
\end{equation}
Let us define functions ${Q_\mu}$ in terms of magnitudes of complex quantities ${\Du{\mu}x_\mu}$
\begin{equation}\label{Qdef2}
Q_{\mu}=2|\Du{\mu}x_{\mu}|^2%=2\Du{\mu}x_{\mu}\,\Db{\mu}x_{\mu}
\comma
\Du{\mu}x_{\mu}=\srh\sqrt{Q_{\mu}}\,e^{i\alpha}
\period
\end{equation}
The orthonormal Darboux basis is not fixed
by conditions \eqref{ghdiag} uniquely. There remains
a freedom of a rotation in each KY 2-plane, which in terms
of the null basis  \eqref{mdef} reads
${\muv{\mu}\to \exp(i\varphi_\mu)\muv{\mu}}$. We uniquely 
fix the Darboux basis by setting the phase factor
${\alpha=\pi/2}$. Then, we have
\begin{equation}\label{eqn1}
\Du{\mu}x_{\mu}=i\srh\,\sqrt{Q_{\mu}}\period
\end{equation}
Using \eqref{Dmx} and \eqref{eqn1} we find
\begin{equation}\label{xiform}
  \PKV = \srh\sum_\mu \sqrt{Q_\mu}\,(\muv{\mu}+\mbv{\mu})=
  \sum_\mu\sqrt{Q_\mu}\,\ehv{\mu}.
\end{equation}

Eqs.~\eqref{Dmx} and \eqref{eqn1} also give us that
the gradient ${\grad x_\mu}$ of the eigenvalue function ${x_\mu}$ is proportional
to ${\enf{\mu}}$,
\begin{equation}\label{dxenf}
  \grad x_\mu = \sqrt{Q_\mu}\,\enf{\mu}\period
\end{equation}
A simple calculation employing Eqs.~\eqref{ghdiag}, \eqref{xiform} 
and \eqref{dxenf} shows that
\begin{equation}\label{xih}
  \PKV\ctr\pKY = -\sum_\mu x_\mu\sqrt{Q_\mu}\,\enf{\mu}
    =\grad\Bigl(-\frac12\sum_\mu x_\mu^2\Bigr)\period
\end{equation}
With the help of the fact that this 1-form is exact and using 
the closeness of ${\pKY}$ we immediately obtain the desired relation
\begin{equation}\label{Lixihqed}
  \lied_\PKV \pKY = \PKV\ctr\grad\pKY + \grad (\PKV\ctr\pKY) = 0\period
\end{equation}

\sect{Killing vector condition for the Einstein spaces}
The second condition in Eq.~\eqref{Liexihg}, which 
states that ${\PKV}$ is a Killing vector, can be easily proved
if the Einstein equations are imposed.
It was demonstrated in \cite{Tachibana} that 
\begin{equation}\label{Snablaxi}
 \nabla_{(a} \xi_{b)} = \frac1{D-2}\,R_{n(a} h_{b)}{}^{n}\period
\end{equation}
 %The proof of this relation involves taking the covariant derivative of the definition \eqref{xidef} of ${\PKV}$, 
%commuting both derivatives obtaining the expressions with 
%the Riemann and consequently Ricci tensors, and finally using symmetries of the Riemann tensor
%and the observation that ${\nabla_n\xi^n=0}$.
For spaces satisfying the Einstein vacuum equations with the cosmological constant
we have the Ricci tensor proportional to the metric 
and thanks to the antisymmetry of ${\pKY}$ we immediately get
${\nabla_{(a}\xi_{b)} = 0}$, that is ${\lied_\PKV \mtrc =0}$.
%\begin{equation}\label{Liexigqed}
%  \nabla_{(a}\xi_{b)} = 0 \quad \Rightarrow \quad \lied_\PKV \mtrc =0\period
%\end{equation} 
Thus, on-shell the conditions \eqref{Liexihg} are valid and using
the results of \cite{HouriEtal} one can derive that the metric represents
the Kerr-NUT-(A)dS spacetime.

\sect{Construction of the canonical form of the metric}
If we do not impose the Einstein equations, 
it is not a straightforward task to prove that ${\PKV}$ is a Killing vector.
Therefore, instead we proceed in a different way---we prove directly the existence of
coordinates ${\psi_j}$ and show that the metric can be written in the canonical form 
\mbox{\eqref{canform}--\eqref{Qdef}}. Here we sketch only main
steps, the details will be discussed in a more technical paper \cite{proofpaper}.

First, taking all projections of equation \eqref{eqn}, 
we collect a partial information about the Ricci coefficients. For example, we obtain
that only those Ricci coefficients with at least two indices equal are nonvanishing.
%\begin{equation}\label{riccic1}
%\begin{gathered}
%  \riccic{\uix\mu}{\uix\mu\uix\nu} = -\riccic{\bix\nu\uix\mu}{\bix\mu} 
%     = -\frac{i}{\sqrt2} \frac{\sqrt{Q_\nu}}{x_\mu-x_\nu}\commae\\
%  \riccic{\uix\mu}{\uix\mu\bix\nu} = -\riccic{\uix\mu}{\uix\nu\bix\mu} 
%     = -\frac{i}{\sqrt2} \frac{\sqrt{Q_\nu}}{x_\mu+x_\nu}\period
%\end{gathered}
%\end{equation}
Next, using ${\PKV\ctr\grad x_\mu=0}$ we can calculate the Lie derivative of ${\enf{\mu}}$
in terms of function ${\hat{q}_\mu=\PKV\ctr\grad(\ln\sqrt{Q_\mu})}$. Using duality relations
and action of the principal CKY tensor we find
\begin{equation}\label{Lieev}
  \lied_\PKV \env{\mu} = \hat{q}_\mu \env{\mu} + \sum_\nu E^{\nu}_{\mu}\,\ehv{\nu}\comma
  \lied_\PKV \ehv{\mu} = -\hat{q}_\mu \ehv{\mu}\commae
\end{equation}
where ${E^{\nu}_{\mu}}$ are yet unspecified components.
Expressing these Lie derivatives using covariant derivatives gives
an additional information about the Ricci coefficients
and determines the components ${E^{\nu}_{\mu}}$ in terms of 
the Ricci coefficients and derivatives of ${Q_\mu}$. 
It also guarantees ${\ehv{\nu}\ctr\grad Q_\mu=0}$ for ${\mu\ne\nu}$
and ${\hat{q}_\mu=\ehv{\mu}\ctr\grad\sqrt{Q_\mu}}$. These facts allow
us to calculate the Lie brackets among all vectors ${\env{\mu},\ehv{\mu}}$ of the Darboux basis.
They do not commute, with the exception of \vague{hatted} ones: ${[\ehv{\mu},\ehv{\nu}]=0}$.

Now, we introduce a new basis ${\{\cxv{\mu},\chv{j}\}}$, ${\mu=1,\dots,n}$, ${j=0,\dots,n-1}$,
\begin{equation}\label{cvframe}
  \cxv{\mu} = \frac1{\sqrt{Q_\mu}}\,\env{\mu}\comma
  \chv{j} = \sum_\mu A^i_\mu\sqrt{Q_\mu}\,\ehv{\mu}\commae
\end{equation}
with ${A^i_\mu}$ given by \eqref{AUdef}.
The geometrical meaning of ${\chv{j}}$ can be elucidated by observing
that ${\chv{j}=\tens{K}_j\ctr\PKV}$, where ${\tens{K}_j}$
is the ${j}$-th Killing tensor in the tower of 2-rank Killing tensors built
from the principal CKY tensor in~\cite{KrtousEtal:2007a}.\nopagebreak

Using the known Ricci coefficients and the Jacobi identity 
we can prove that vectors of this frame do commute,
\begin{equation}\label{LieBrcframe}
  [\cxv{\mu},\cxv{\nu}]=[\cxv{\mu},\chv{j}]=[\chv{i},\chv{j}]=0\period
\end{equation}
Moreover, for the dual frame 
\begin{equation}\label{cfframe}
  \cxf{\mu} = \sqrt{Q_\mu}\, \enf{\mu} = \grad x_\mu\comma
  \chf{i} = \sum_\mu \frac{(-x_\mu^2)^{n{-}1{-}i}}{U_\mu\sqrt{Q_\mu}}\,\ehf{\mu}
\end{equation}
we show
\begin{equation}\label{gradcframe}
  \grad\cxf{\mu}=0 \comma \grad \chf{\mu} = 0\period
\end{equation}
Both conditions \eqref{LieBrcframe} and \eqref{gradcframe}
ensure that additionally to ${x_\mu}$, ${\mu=1,\dots,n}$,
it is possible to introduce coordinates ${\psi_j}$, ${j=0,\dots,n-1}$,
such that
\begin{equation}\label{coors}
\begin{gathered}
  \cxv{\mu}=\cv{x_\mu}\comma\! \chv{i} = \cv{\psi_i}\;\,\text{and}\;\,
  \cxf{\mu}=\grad x_\mu\comma\! \chf{i} = \grad \psi_i\period
\end{gathered}
\end{equation}
Taking into account the inverse of Eqs.~\eqref{cfframe} we get
\begin{equation}\label{efcoor}
  \enf{\mu}=\frac1{\sqrt{Q_\mu}}\,\grad x_\mu\comma
  \ehf{\mu}=\sqrt{Q_\mu}\,\sum_{i=0}^{n-1}A^i_\mu\, \grad\psi_i\period
\end{equation}
Substituting \eqref{efcoor} into \eqref{ghdiag}
leads to the metric \eqref{canform} with unspecified metric functions ${Q_\mu}$.
However, in the process, we also learn that metric functions ${Q_\mu}$
must take the form \eqref{Qdef}, particularly that ${\hat{q}_\mu=0}$ and ${E^{\nu}_{\mu}=0}$. 
This finishes the proof of our main result:
we have constructed a coordinate system 
in which the off-shell metric takes the canonical form \eqref{canform}--\eqref{Qdef},
starting only from the quantities determined by the principal CKY tensor.

As a consequence we have also established that
${\PKV}$ is a Killing vector which we call the \emph{primary} one.
Thus we proved both conditions \eqref{Liexihg} without employing 
the Einstein equations. Actually, all vectors ${\cv{\psi_j}}$ are Killing 
vectors---obtained by the action of the Killing tensors $\tens{K}_j$ 
on the primary Killing vector ${\PKV=\cv{\psi_0}}$.
They coincide with vectors used in \cite{HouriEtal}.

\sect{Summary}
As we have already mentioned, several important results were earlier obtained
for the general off-shell metric in a canonical form
\eqref{canform}-\eqref{Qdef}. Namely, this metric is of the type D 
\cite{HamamotoEtal:2007,PravdaEtal:2007}.
It allows separation of variables for the Hamilton--Jacobi,
Klein--Gordon and Dirac equations
\cite{FrolovEtal:2007,OotaYasui:2007}. 
Geodesic equations are completely integrable \cite{PageEtal}
and there exists a complete set of integrals of motion
which are linear and quadratic in momenta \cite{KrtousEtal:2007a}. 
Since the canonical form of the
metric follows from the existence of the principal CKY tensor, all
these properties are common for spacetimes which admit such a tensor.

In our consideration we have focused on a generic case when the principal
CKY tensor is non-degenerate. In a degenerate case some of the
eigenspaces of $\pKY$ may have more than 2 dimensions. We have also
focused on an Euclidean form of the metric. After the Wick's
rotation which transforms the Euclidean metric to the Lorentzian one,
it may happen that some of the coordinates $x_{\mu}$ become
null. Additional degeneracy may be created when the
primary Killing vector $\PKV$ vanishes. 
All these special degenerate cases require additional consideration. 

\sect{Acknowledgments}
P.K.\ is supported by  the grants GA\v{C}R 202/08/0187, M\v{S}MT~\v{C}R LC06014, 
and appreciates the hospitality of the University of Alberta. 
V.F.\ thanks the Natural Sciences and Engineering Research 
Council of Canada and the Killam Trust for financial support. 
D.K.\ is grateful to the Golden Bell Jar Graduate
Scholarship in Physics at the University of Alberta. 
We have benefited from discussions with Don N. Page.

%\bibliographystyle{H:/odborne/library/TeX/bst/longprsty}
%\bibliography{H:/odborne/library/TeX/bib/references}
%\end{document}

%%%%%%%%%%%%%%%%%%%%%%%%%%%%%%%%%%%%%%%%%%%%%%%%%%%%%%%%%%%
%% references for arXive

\end{document}